# Improving Interaction with Virtual Globes through Spatial Thinking: Helping Users Ask "Why?"

**Johannes Schöning**
Institute for Geoinformatics
University of Münster
Robert-Koch-Str. 26-28
48149 Münster, Germany
j.schoening@uni-muenster.de

**Brent Hecht**
Department of Geography
University of California, Santa Barbara
Santa Barbara, CA 93106
United States
bhecht@geog.ucsb.edu

**Martin Raubal**
Department of Geography
University of California, Santa Barbara
Santa Barbara, CA 93106
United States
raubal@geog.ucsb.edu

**Antonio Krüger**
Institute for Geoinformatics
University of Münster
Robert-Koch-Str. 26-28
48149 Münster, Germany
antonio.krueger@uni-muenster.de

**Meredith Marsh**
Department of Geography
University of California, Santa Barbara
Santa Barbara, CA 93106
United States
meri@geog.ucsb.edu

**Michael Rohs**
Deutsche Telekom Laboratories
TU Berlin
Ernst-Reuter-Platz 7
10587 Berlin, Germany
michael.rohs@telekom.de

**ABSTRACT**
Virtual globes have progressed from little-known technology to broadly popular software in a mere few years. We investigated this phenomenon through a survey and discovered that, while virtual globes are *en vogue*, their use is restricted to a small set of tasks so simple that they do not involve any spatial thinking. Spatial thinking requires that users ask "what is where" and "why"; the most common virtual globe tasks only include the "what". Based on the results of this survey, we have developed a multi-touch virtual globe derived from an adapted virtual globe paradigm designed to widen the potential uses of the technology by helping its users to inquire about both the "what is where" and "why" of spatial distribution. We do not seek to provide users with full GIS (geographic information system) functionality, but rather we aim to facilitate the asking and answering of simple "why" questions about general topics that appeal to a wide virtual globe user base.

**ACM Classification Keywords**
H.5.2 [Information Interfaces and Presentation]: User Interfaces

**INTRODUCTION**
There exists myriad evidence of the dramatic rise in popularity of virtual globes. Google Earth [23], the most ubiquitous virtual globe, was downloaded over 100 million times in its first 15 months [39] of release. U.S. President George W. Bush has said that he uses Google Earth to look at his Texas ranch [22]. Moreover, the phenomenon has even inspired a *Nature* news feature [10].

The *Nature* article notes an important dichotomy between the features employed by the casual user of Google Earth and those used by the scientific audience. The author writes "to the casual user ... the appeal of Google Earth is the ease with which you can zoom from space right down to the street level" while the attraction of scientists and enthusiasts to the program lies in the fact that it is "an easy way into GIS software" (p. 776). While virtual globes' use as an entryway into the world of GIS cannot be understated, this dichotomy raises doubts about the ground-breaking nature of the technology on the large group of people who do not make the jump to advanced GIS packages. The results of a survey, discussed later in the paper, elicit further concerns about the superficiality of tasks performed by the average virtual globe user.

As defined in the recently published National Research Council Report, Learning to Think Spatially [12] spatial thinking (in the geospatial domain) is a "dynamic process that allows us to describe, explain, and predict the structure and functions of objects and their relationships in real and imagined spatial worlds." (p. 33) A significant part of the spatial thinking process involves generation of hypotheses, pattern predictions, and tests of hypotheses. Essentially, when thinking spatially, individuals observe what patterns exist in the environment and seek to provide explanations for these patterns. In short, these individuals ask "what is where?" and "why?".

GIS is increasingly heralded as the most probable support system for facilitating the spatial thinking process as it allows for spatialization [17] of non-spatial datasets. The spatial representation of data permits the individual to ask "why" questions – i.e. why certain patterns or relationships exist in or between certain places – questions that are difficult to formulate when the same data is experienced in a different format (e.g. a spreadsheet). With expertise in traditional GIS technology, these patterns and processes can be further explored using spatial statistics and other advanced operations, analyses certainly beyond the knowledge of the everyday Google Earth user.

As virtual globe technologies become increasingly pervasive, much hope surrounds their capacity to potentially enhance spatial thinking ability among both K-12 students and non-expert users. However, as demonstrated in the results of a survey on the uses of virtual globes (see below), most individuals use these technologies simply for observational purposes, and little to no spatial thinking actually occurs. In other words, the majority of individuals seem to use these technologies to observe the "what" of spatial data (e.g. the location of their home or business and where it is in relation to other prominent geographic features), but moving beyond pure observation to questioning why certain patterns exist in the landscape proves out of reach to the casual user. As typical virtual globe technologies are not coupled with specific datasets or feature sets, and adding data to the existing software involves a certain level of expertise, the majority of individuals does not have access to the information or tools they need to ask the "why" questions. Therefore, the technologies do not, in their current form, typically support the spatial thinking process.

Importantly, this research is not an effort to incorporate an easy-to-use GIS into a virtual globe software package. Other projects such as ArcGIS Explorer [16], Google Earth [23] itself, and Mapalester [25] have tackled this problem to at least a small extent. Our aim is entirely different. Rather than providing the user with advanced GIS functionality (e.g. spatial join, cluster analysis, buffers) to answer spatial thinking questions, our system facilitates the asking and answering of *simple* "why" questions, e.g. Why does this spatial feature display this value? What is the relationship between these two features?

Our prototype (see figure 1) enables this facilitation by demonstrating enhancements in two key areas: *data type* and *interface*. We introduce a new simple spatial thinking-oriented virtual globe data type called Explicitly Explanatory Spatial Data (EESD), which contains both a standard spatial layer and a new *explicitly explanatory* layer designed specifically to answer "why" questions. Two test case data sets are presented. The first is based on our previous WikEar [38] and Minotour [27] projects, which use Wikipedia to generate narratives between geotagged Wikipedia articles. The second uses a prototype of GeoSR [26], a new semantic relatedness-based system that is backed by a Wikipedia-based knowledge repository. The semantic relatedness literature originates in computational linguistics and seeks to define a single number to "quantity the degree to which [any] two concepts are related" [4, p. 1].

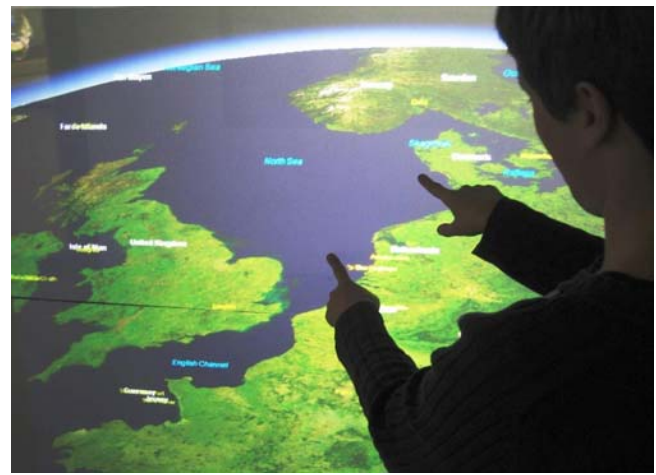

**Figure 1. Usage of our virtual globe prototype on a multi-touch surface (user is zooming).**

The following section places this paper in the context of the related work in the variety of fields that provide the basis for this research. The third section covers the analysis and results of a survey conducted on virtual globe use in Münster, Germany. In the following section, we describe the conceptual design of our system from both a data type and interface perspective. Our implementation is discussed in the fifth section and the sixth section contains a presentation and discussion of the results of an informal evaluation of the interaction with our new system. We conclude with a discussion of future work.

**RELATED WORK**
NASA World Wind [35] is the second biggest player in the virtual globes market behind Google Earth [23]. While

Google Earth is targeted at a general audience, NASA World Wind can be more easily customized for specialized groups of users. As a browser plug-in, Microsoft's Windows Live Local Earth 3D [32] also provides an interface to a variety of high-resolution satellite images and maps. ArcGIS Explorer [16] from ESRI, the leader in the professional GIS market, is a client for ArcGIS Server and supports WMS (Web Map Service).

The two prototype EESD layers developed in this research draw from a variety of disciplines. The first layer, which is based on Minotour and WikEar, is rooted in the field of intelligent narrative technologies (INT). Although it is unique in its approach on the technical side, it is firmly based on the narrative theory developed in [28], [6] and others. The second layer is motivated by previous semantic relatedness measures, such as those described in [8]. Three other relatedness measures based on the Wikipedia corpus have been published: WikiRelate! [40], Explicit Semantic Analysis [20], and the work of Zesch et al. [42]

Critically, our new wall-size interface – based on multi-touch technology – is the facilitator between the "what is where" and "why" questions and the new data type, and is designed to maximize ease of use. Our use of multi-touch was inspired by a desire to take advantage of advances in technology to optimally assist the user in posing and receiving answers to simple spatial thinking questions.

We use a low-cost, large-scale (1.8 x 2.2 meter) multi-touch surface that utilizes the principles of frustrated total internal reflection (FTIR), greatly increasing the near-future practicality of our prototype because such interfaces are cheap and quick to build. Jeff Han presented the original FTIR multi-touch sensing work in February 2006 at the Technology Entertainment Design (TED) Conference [24]. Total internal reflection is an "optical phenomenon that occurs when a ray of light strikes a medium boundary at an angle larger than the critical angle with respect to (a) normal to the surface" [43]. Changing the refraction index by touching the medium effectively creates an infrared light spot under the touched area. Detecting this spot with a camera behind the multi-touch surface and applying simple computer vision algorithms to calculate the position of the touch on the surface is straightforward. These surfaces, capable of sensing fingers, hands, and even whole arms, can be constructed from readily available components. That said, the steps involved in building a high-quality FTIR-enabled surface on both a software and hardware level are not trivial and require much engineering effort. If multi-touch applications need to distinguish between different users the "Diamond Touch" [15] concept from MERL could be used, with the drawback that the users either need to be wired or stay in specially prepared locations. Because it is less important in our work to distinguish between different users, we determined that the benefits of using FTIR far outweigh the disadvantages.

The selection of relevant data, the configuration of adequate data presentation techniques, and the input or manipulation of data are central tasks in an interactive system. A criticism of many previous multi-touch projects is that the model of interaction does not change at all from previous interaction paradigms. In Han [24], the iPhone [2], Microsoft Surface [33], etc., two-finger gestures are the only interaction paradigm in which the capabilities of multi-touch are used. We make an effort to use the full potential of multi-touch in providing the user an intuitive way to interact with our prototype.

With more and more technology being embedded into the environment, new interaction paradigms that go beyond the traditional WIMP metaphor have arisen, several of which are relevant to our work. Virtual globes still require displays, but these can be of arbitrary sizes. Larger displays (100 cm and more) are especially suited to our virtual globe prototype. Mice and keyboards can be used to navigate a virtual globe, but are not optimal devices for this purpose (e.g. special 3D-space mice [1] do exist to operate Google Earth efficiently). Multi-touch ([5] and [18]) has been shown to work well with large screens due to its support of multi-finger and bi-manual operation [11].

**SURVEY**

A user survey was conducted to investigate the usage and user needs of virtual globes. The study included 120 participants: 60 female and 60 male. They were randomly selected in a pedestrian area in Münster, Germany and had a mean age of 34.2 years (SD=8.7). The length of the survey was about 5 to 8 minutes, during which time each participant was asked ten questions about her or his knowledge and use of virtual globes, as well as digital maps. We also asked about digital maps to investigate the usage similarities and differences between the two geovisualization mediums.

First, the participants were asked if they were aware of digital maps; 89.2% (± 5.5%) of the participants answered in the affirmative, of whom 92.5% (82.5% ± 6.8% overall) use digital maps more than 5 times per month. When asked identical questions about virtual globes, 67.5% (± 8.3%) said that they were aware of virtual globes while 59.2% overall (± 8.8%) said they used them more than 5 times per month.

We then asked users about the motivations behind their virtual globe and digital map use. Around half (53.4% ± 11.6%) said they used virtual globes for either looking at their own house or other individual places (e.g. a neighbor's house, their hotel from their last vacation, the city center). The second most common uses of virtual globes were navigation (16.9% ± 8.7%) and locating businesses (14.1% ± 8.1%). More esoteric responses, such as that of a roofer who said he used Google Earth to find roofs that needed repair, rounded out the respondents' uses. The distribution of digital map use was quite different than that of their virtual globe cousins, with over 50% of respondents saying that they used digital maps for navigation. More details on both results can be seen in figures 2 and 3.

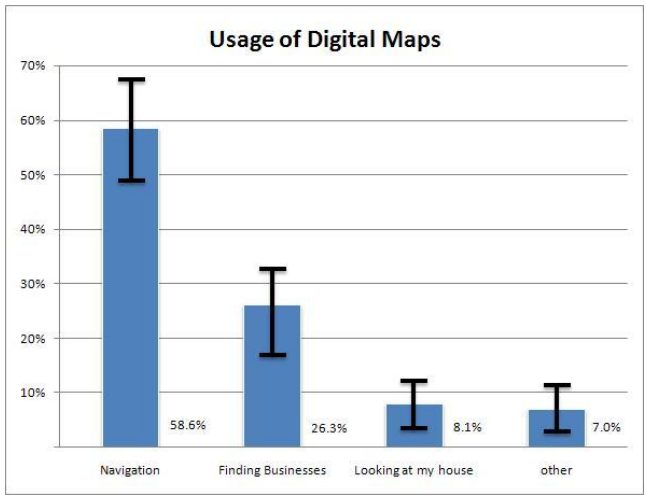

Figure 2. Usage of Digital Maps.

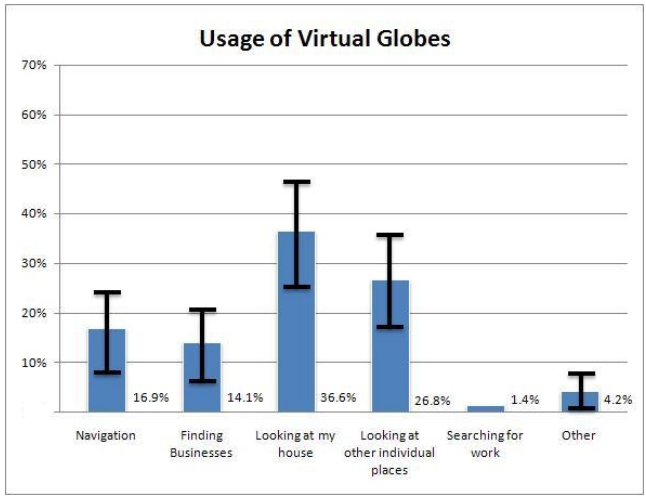

Figure 3. Usage of Virtual Globes.

Finally, we asked respondents to compare and contrast the advantages of virtual globes and maps. They answered that the main advantages of digital maps were easy navigation and global map coverage. In contrast, the main advantages of virtual globes over digital maps were the ability to view high resolution satellite images and aerial photography overlaid with additional information such as geotagged Wikipedia articles and Panoramio [36] photos in a 3D environment.

**DESIGN REQUIREMENTS**
As noted in the introduction, the central design conclusion of the virtual globe survey is that the majority of tasks employed by virtual globe users are simplistic and do not require spatial thinking. As spatial thinking involves both noticing patterns in the landscape (whether it be a real or represented environment), *and* questioning the evolution of those patterns, simple observational activities do not constitute spatial thinking. Similarly, with navigation and business location (in this case), virtual globes are employed simply to answer the "what" question, as well as "where" certain features are in relation to one another. There is no "why" in the picture.

It is also important to draw conclusions – albeit less firm ones – from trends that can be found in the unstructured and unsolicited responses from survey participants. First and foremost, users like the general idea of displaying the Earth in three dimensions, as they indicated they enjoyed viewing the Earth as it truly is. However, they noted that the interaction with a 3D environment was difficult and many expressed a desire for easier-to-control interfaces [7]. Finally, and most critically, over the half of the users indicated that they felt that virtual globes could be more useful to them if they only could figure out more tasks to perform with them (besides those they indicated). This can be interpreted as a desire to engage in more advanced tasks, likely those involving some spatial thinking.

**CONCEPTUAL DESIGN**
Following the results of our survey, we developed a new virtual globe prototype designed to widen the potential uses of the technology by allowing users to spatially inquire about both "what" *and* "why".

**Data**
There were two key challenges in developing the data type for our improved virtual globe prototype. The first was to design a general data structure that would enable users to both ask and answer spatial thinking questions. The second was to appeal to the thematic interests of the broad virtual globe user base. The former challenge is the topic of the first subsection and the latter is discussed in the second.

*A Framework To Facilitate Answering "Why" Questions of Data*
GIS software for years has enabled users to engage in a large variety of advanced spatial thinking tasks. However, the design goal for this research is to facilitate simple versions of these tasks using intuitive paradigms in virtual globes. Our solution on the data side is the Explicitly Explanatory Spatial Data (EESD) type. Each EESD set is defined by two layers. The first layer is the standard spatial data layer that has been in use since the first GIS around 40 years ago. It can contain raster cells, points, polylines, polygons, or any other feature type that can be displayed on a virtual globe. This layer – in an abstract sense, at least – also contains attribute data for the features. The second layer, the explicitly explanatory layer, holds the innovation. This layer contains *explicit* explanations for the attribute values and/or relationships present in the spatial data layer. It is hypothesized that explanation of these two properties of a spatial data layer, corresponding to the "objects" and "relationships" noted in the definition of spatial thinking found in the introduction, will best facilitate basic spatial thinking tasks. This layer must make it a trivial matter for the interface – responding to a "why" query from the user – to return an explanation.

*Examples of EESD Sets*

We have implemented two examples of the EESD sets, the WikEar [38] data set, which is derived from our Minotour [27] work, and the data set generated by an early version of our GeoSR [26] project. Both have a spatial data layer that is generated from the large number of hand-geotagged articles in the English version of Wikipedia. The former EESD set is of the type that contains explanatory information about spatial relationships while the latter is focused on explaining single data values (although users will likely identify implicitly explained patterns as well).

Before detailing the prototype EESD sets, however, it is important to discuss certain properties of the Wikipedia knowledge repository. First and foremost, it is necessary to acknowledge concerns about the risks of using Wikipedia data. Denning et al. [13] codified these risks into concerns over accuracy, uncertain expertise, volatility, coverage, and sources. However, Giles [21] reported that Wikipedia is comparable to the Encyclopedia Britannica in terms of number of serious errors and only slightly worse than Britannica when it comes to "factual errors, omissions, or misleading statements". Regardless, given the requirements of this research: a natural language knowledge repository with both an extensive and intensive coverage of world knowledge, Wikipedia is by far the best choice. With over 2 million articles in the English version (as of submission) and 14 other language editions with over 100,000 articles (all methods described here work with all Wikipedia languages), Wikipedia is the largest Encyclopedia to ever exist.

For the purposes of this research, Wikipedia articles can be split up into 3 groups: (1) articles without a geotag, (2) articles with a geotag, and (3) articles about purely temporal phenomena (i.e. the article on the year "1983" or the date "October 1"). We call articles in the first group "non-spatial articles" and articles in the second "spatial articles". The third group exists because purely temporal articles have very defined relationships encoded in their links with other articles; linking to a temporal article is nothing more than providing an explicit temporal reference to the article, something that can be useful in some contexts but amounts to enormous noise in this work.

Finally, the concept of a Wikipedia "snippet" is critical to both EESD sets. In a general sense, a Wikipedia snippet is simply a paragraph of a Wikipedia article. These paragraphs are unique in natural language text knowledge repositories in that they are almost entirely independent of one another. In other words, snippets almost never contain unexplained or incomplete textual references to other snippets. This is a direct result of the encyclopedic writing style that is the Wikipedia norm, as well as the collaborative nature of Wikipedia, in which the median number of authors per article (as of 2006) in the English version was over seven [9]. We have found experimentally that the only context necessary for fully understanding the vast majority of snippets is the title of the article to which the snippet belongs. One can further increase understanding of independent snippets by providing the hierarchy of headings in which the snippet resides (i.e. for the United States article, there are 3 snippets under "History->Native Americans and European Settlers" as of November 26, 2007).

Minotour [27] generates cohesive stories from a Wikipedia knowledge repository using a data mining methodology derived from narrative theory. The WikEar [38] dataset contains human-narrated versions of Minotour's stories in an attempt to simulate future text-to-speech technology. The stories begin at one Wikipedia article $a$, end at a Wikipedia article $b$, and contain $s$ snippets, each of which belong to a Wikipedia article on a narrative-theory defined optimal path from $a$ to $b$ through the Wikipedia Article Graph (WAG). In the WAG, each article is a vertex and each directional link between articles is an edge. The variables $a$, $b$, and $s$ are all user-defined.

The primary test case for Minotour and WikEar is the generation of educational tourism narratives. In this context, spatial Wikipedia articles are used for $a$ and $b$, while non-spatial articles provide the snippets for the body of the narrative. Critically, applied in this manner, Minotour narratives, by definition, explain a relationship between the spatial entities that articles $a$ and $b$ describe. As such, operating with a layer of the spatial references of Wikipedia articles, the narratives form an explicitly explanatory layer for the relationships between the points in the spatial layer. Looking at the spatial layer, a user can ask, "Why are these two spatial entities related?" and the system can easily respond with an answer. With tens of thousands of spatial articles in the English Wikipedia, users are able to ask this very simple and general spatial thinking question about almost anywhere in the world. This simplicity and generality fits in with other typical virtual globe data layers (i.e. satellite photography), but also allows for the explicit answering of "why" questions.

An early prototype of GeoSR is the backbone of the second EESD layer. GeoSR is based on our novel ExploSR semantic relatedness (SR) measure, the first adapted to the context of data exploration. The goal of SR measures is to identify a value that summarizes the number of relationships between two entities as well as the strength of these relationships. By analyzing the Wikipedia Article Graph (WAG), ExploSR derives such values between the entities represented by Wikipedia articles. The key variables looked at by ExploSR when examining any two articles $a$ and $b$ are the myriad paths from $a$ to $b$ (and vice versa) in the WAG and the scaled weight of the links in those paths. Link weights are determined by a mixture of article length, number of out-links (outdegree) between the linked articles, text position of those links, and several Wikipedia-specific variables.

In addition to being the first semantic relatedness measure designed for use in a data exploration context, ExploSR is the first measure to utilize the WAG and the first measure that can be visualized in a reference system (in this case, a geographic one) [26]. Further discussion of the benefits of the WAG for this type of semantic relatedness application is merited. First, the WAG is ideal for SR measures designed with data explanations in mind because a natural language explanation is built into every outputted measure (see below). Secondly, the WAG is replete with both classical relationships, i.e. *is-a* (hypernymy and hyponymy) and *has-a* (meronymy and holonymy), and non-classical relationships [34]. We have found qualitatively that these non-classical relationships such as "spoke-at", "ate-a", "wrote-about", "tool-he-uses-to-look-at-ranch" to be far more important than their more standard cousins when evaluating SR measures on articles representing entities that belong to a commonly-used reference system, such as spatial and temporal articles.

Abstractly, ExploSR takes a Wikipedia article as an input and returns a single semantic relatedness value from $a$ to all Wikipedia articles of type $T$. Users can then query GeoSR for an explanation of any value, and GeoSR will return the snippets containing the links that form the path between $a$ and $b$ in the WAG, where $b \in T$. Typically, $T$ will be a set of articles that all belong to some semantic reference system [30, 31] (i.e. spatial or temporal).

There are many geographic applications of GeoSR, two of which are used in our GeoSR-derived prototype EESD data set. The first occurs when $a$ is a non-spatial article and $T$ equals the set of spatial articles. This will result in all spatial articles having a semantic relatedness value to $a$. The second occurs when $T$ again equals the set of spatial articles, but $a$ is also a spatial article. While similar, these applications differ significantly in that one result in measures of theme-to-spatial entity relationships while the other outputs spatial entity-to-spatial entity relationships. Both applications, however, provide excellent EESD sets. In both, the spatial data layer is a spatial visualization of the GeoSR measure in which the spatial entities depicted are those about which there are Wikipedia articles. In our example (see figures 5, right, and 6), this layer is represented as a graduated symbol map, with each spatial Wikipedia article depicted as a point in a geographic reference system. Other visualizations are also possible with improved georeferencing of Wikipedia articles (for instance, referencing articles about spatial entities of sufficiently large extent to polygons rather than points). The size of the symbol is defined by the value of the semantic relatedness measure, with bigger symbols indicating more and/or stronger relationships. The explanatory layer, is built directly into the ExploSR system as described above. Applied in a geographic context, this amounts to every value visualized on the map having an explicit explanation found easily in the data set.

Similarly to the WikEar dataset, the GeoSR system generates data of broad, general appeal. Since $a$ can be any article, the user is able to see how related all entities described by Wikipedia articles that are spatially referenced are to any entity in all of Wikipedia, from "multi-touch" to "George W. Bush" to "Rugby" to "Surfing". Importantly, both layers are easily applied to spatial subsets, or "extents", of the globe. Using the measures on small extents will focus the graduated symbol visualization to allow maximum differentiation in relatedness in the region of study.

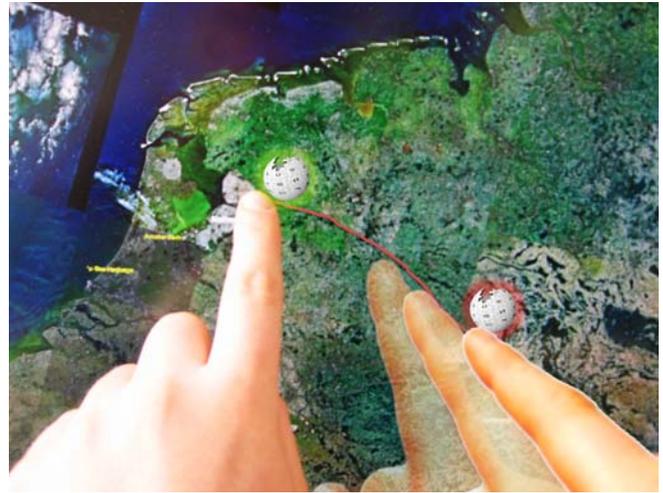

**Figure 4: Interaction with the first (Wikear) EESD Layer.**

**Interaction with the data**
To benefit from the EESD layers, users need an intuitive way of interacting with them. To provide this intuitive interaction paradigm, we use the full advantages of our multi-touch surface.

The basic spatial interaction tasks such as pan, rotate, zoom and tilt are implemented using the principles shown in the video by Han [24] and the iPhone [2]. For instance, the user can pan through the world with the flick of a finger or hand and use other multi-touch gestures to zoom, rotate, tilt or navigate. "Click" and "double-click" are implemented with simple taps.

*Interaction with the First (WikEar) EESD Layer*
The interaction with the first EESD layer is straightforward. The user selects two spatial Wikipedia features by double-clicking (double-touching) Wikipedia icons (which indicate spatial Wikipedia articles) simultaneously with two fingers (see figure 4). The icon selected by the one hand is the start feature $a$ and the other Wikipedia feature is the end feature $b$. The start feature, end feature and a line between them are highlighted and a story derived from Wikipedia (as described in the previous section) is read out to the user. Users can control the speed of playback by dragging their finger from the start point (in green) to the end point. By moving the finger from the end location to the start location story is derived by swapping the start and end point and th is stories is played back. By releasing her/his fingers from the multi-touch surface, a user can stop playback and can, for example, navigate to another place on the earth or request other information.

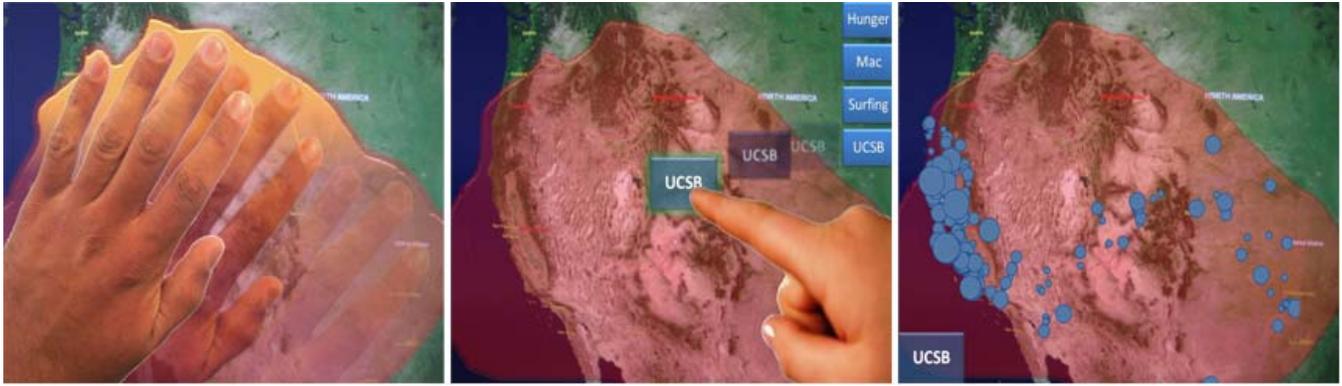

**Figure 5: Interaction with the second EESD Layer – Region Selection (left). Interaction with the second EESD Layer – Dropping a theme into a region (middle). Interaction with the second EESD Layer – Visualization of the result (right).**

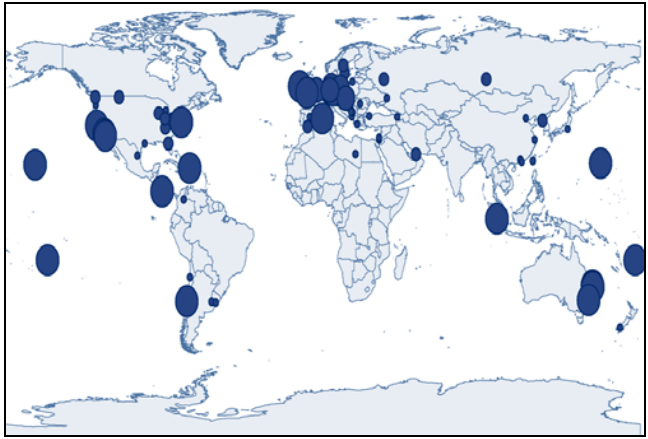

**Figure 6: Visualization of the theme "surfing" (second EESD layer) on a globe scale derived from the German Wikipedia.**

*Interaction with the Second (ExploSR) EESD Layer*
To interact with the second EESD layer, users must first define a region. This can be done by activating the "region definition mode" by touching a button and dragging the hand(s) or finger(s) over the multi-touch surface (see figure 5, left). After lifting her/his hand or hands from the multi-touch surface, a user sees a menu where she or he can select one or more different "themes" (which represent different articles *a* as input) for that region (in our prototype we have 25 pre-computed themes). By dragging a theme into the region (see figure 5, middle), users can explore semantic relatedness values for that "theme" in the region they selected (see figure 5, right). Clicking on a single symbol will provide the text-based explanation of the "why" of each value as described in the data section. Without dragging the "theme" into a predefined region, users can explore the relatedness of that "theme" at a global scale (see figure 6). This mode can be deactivated by disabling the EESD view.

**IMPLEMENTATION**
Both EESD layers operate from a significantly pre-processed version of the Wikipedia knowledge repository. The pre-processing takes as input one of the semi-regularly exported "database backup dump files" from Wikipedia. Currently only the English, German, and Spanish files are supported, but with the help of a translator it would be an easy matter to add support for any language version of Wikipedia. For the larger Wikipedias such as English and German, the size of these dump files is remarkable. The latest English dump file as of November 2007, for instance, was about 12.7GB of text. During the pre-processing stage, the dump file is parsed in a Java parsing engine to isolate article, snippet, link structure, spatial, and temporal information, which is then stored in a MySQL database in a variety of tables. A Java API to this database, which is named WikAPIdia [26] is then used by the systems that generate both EESD layers. The API provides basic access to Wikipedia data as well as more advanced graph mining and spatiotemporal features, which are used by both Minotour and GeoSR.

As noted above, our interface is rooted in a 1.8m x 2.2m FTIR-based multi-touch wall. This wall consists of a 12mm thick acrylic plate, in which every four centimeters a hole for an infrared LED was drilled. The acrylic plate was mounted onto a wall and a wide-angle lens digital video camera (PointGrey Dragonfly2 [37]) equipped with a matching infrared band-pass filter was mounted orthogonally at a two-meter distance. As a projection screen, very inexpensive drafting paper was used. For the projector we used a Panasonic PT-AE1000E HD Beamer.

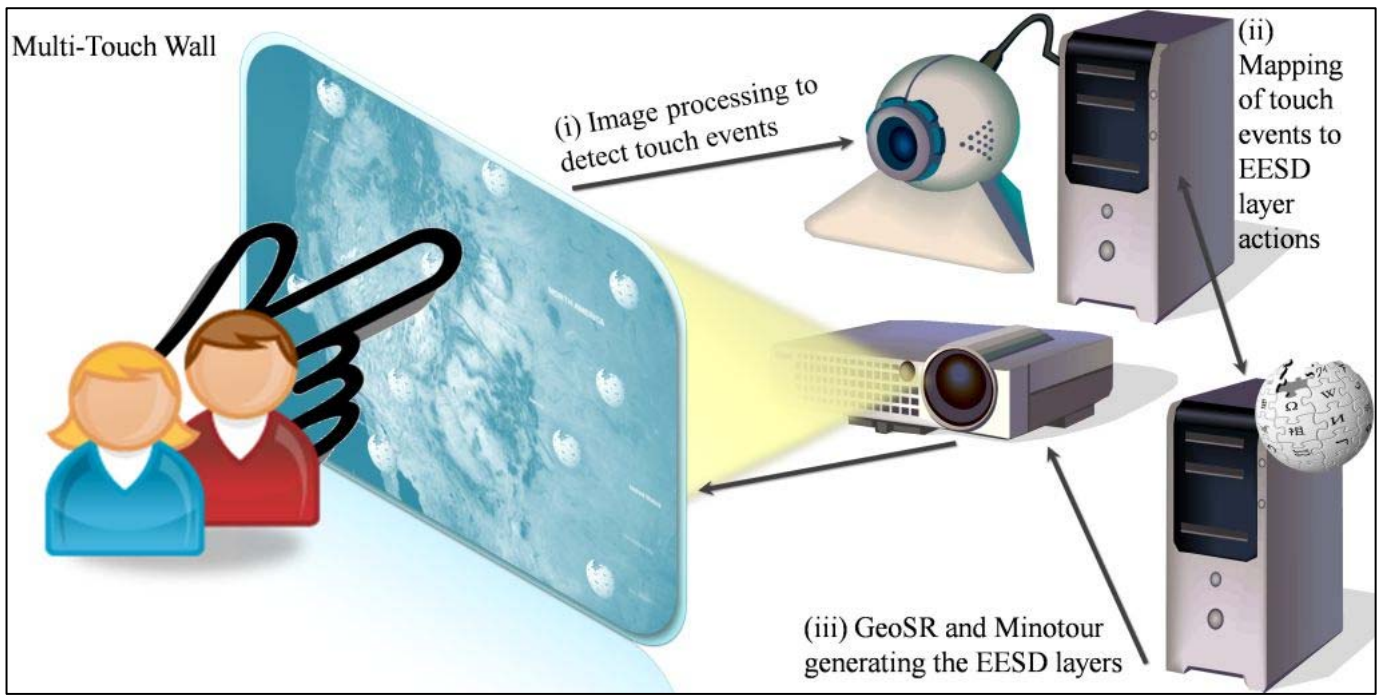

**Figure 7: System overview of the prototype. The virtual globe is based on NASA's World Wind [33] extended with EESD layers and uses the Java-based Multi-Touch Library [14] developed at the Deutsche Telekom Laboratories.**

To improve dragging operations, we placed a thin layer of silicon (Silka Clear 40) between the acrylic and the drafting paper.

The Java-based Multi-Touch Library [14] developed at the Deutsche Telekom Laboratories and released under the GNU Public License was used for image processing. It contains a set of common algorithms designed to work with any multi-touch system such as routines to label connected components and track features. By using an application layer, it is easy to manipulate objects and transform (position, rotate, scale) them. The library also comes with a module for accessing cameras such as the PointGrey Dragonfly2.

Our virtual globe is based on NASA's World Wind [35]. According to [29], NASA World Wind has only one goal: to provide the maximum opportunity for geospatial information to be experienced, regardless of whether the context is education, science, research, business, or government. The NASA World Wind visualization platform is open source and comes with a rich SDK for data set and interface customization, which we take advantage of with our EESD layers and multi-touch interaction.

### State of the Implementation
Currently, we have working prototypes of the EESD layers and an interactive version of our multi-touch virtual globe. To improve the experience of our interface we have to increase the speed of the EESD implementation, further develop the integration between the globe and the data, and improve visualization of the spatial data layer.

### "Sanity Check" of our interaction
Twelve randomly selected employees (9 male, 3 female) of the Institute for Geoinformatics in Münster, Germany (no one who was involved in the project was included) were asked to provide feedback on the interaction with the first and second EESD layers.

Due to the fact that our prototype is not running in real time and that interaction with such an "unready" interface would be distracting to the users, we decided to run an evaluation on a paper mock-up. In doing so, the user acquired "a greater understanding of how the final product will function and the way it will 'look and feel'" [44].

After explaining the possibilities of FTIR multi touch surfaces, the users were asked three questions:

a) How would you choose two spatial features out of a group of features and establish a connection between them? (How to interact with the first layer)

b) How would you select an area as needed in our interaction with the second EESD?

c) How would you assign an attribute from a list to an area?

The answers to these questions were as follows:

a) Eight of the 12 participants would select two features just by clicking (single-touching) the features' icons simultaneously with two fingers, just as we have implemented in our prototype. For icons a small distance apart, one participant would use two fingers of the same hand and for icons further apart one finger of both hands. The four others would double-click (one just single click)

b) Ten of 12 participants would select a region by circumscribing the area with one finger. One of these ten, a trained geographer, would use two fingers simultaneously. Two would use their whole hand to select the region as is established in our prototype. (We let users define a region by using their fingers or of using their complete hand).

c) Seven participants would touch the desired "theme" in the list and then touch inside the selected area. One of these seven users would perform these tasks all at once. The other five participants would drag the theme into the layers as is done in the prototype. When the seven participants who preferred the click interaction were told of the drag method, six of them agreed that the dragging methodology would be more fun and maybe more attractive for an interactive digital globe.

These initial results suggest that while we have to adapt some of our interaction with the virtual globe, most of our interaction paradigm is very intuitive. The informal study also convinced us that users still think in WIMP interaction styles. People have to use multi-touch surfaces more often to become accustomed to their possibilities. After improving the speed of the algorithms we want to formally evaluate the interaction with both EESD layers

**CONCLUSION AND FUTURE WORK**

A common theme the authors' previous collaborative work has been to bring to users of state-of-the-art consumer spatial technologies a fuller sense of knowledge about the world. Too often the gift of spatial context provided by these technologies is under-used by applications that only provide obvious functionality and ignore the users' inclination to explore and learn. For location-based services on mobile devices, this obvious functionality often involves pointing users to the nearest pizza parlor or pub. The corollary for virtual globes is, according to our survey, navigation, sightseeing, and, mainly, innocuous voyeurism. While these applications are certainly useful, much is lost, particularly with respect to the "objects and their relationships" that make up the inspiration for and the answer to spatial thinking questions.

These thus-far missed opportunities do the greatest harm to geography education – both intentional and incidental – something that is widely recognized as severely lacking in many parts of the world. For virtual globes, geography education represents both a largely untapped financial market and a chance to enhance world knowledge. The possibility of virtual globes facilitating the spatial thinking process provides an exciting avenue for this technology to reach its full potential.

We have provided a glimpse of what is possible when spatial thinking is enabled in a virtual globe. However, there is an extensive amount of future work yet to be done. First and foremost, more research must be completed into the current state of virtual globe use. A wider and more structured survey would be extremely useful and could be used to formally derive a requirements analysis for a spatial thinking-facilitating virtual globe prototype. Secondly, more robust theoretical framework for the Explicitly Explanatory Spatial Data (EESD) layers must be developed. Additionally, we must complete formal studies of the interaction with EESD data layers, particularly with regard to the degree to which it enhances spatial thinking.

Separately, much work is being done to develop the EESD types used as prototypes in this work. For instance, implementation speed must be improved. Depending on the entity, the calculation of the second EESD layer can take up to 10 minutes using the Wikipedia data set. Additionally, cartographic research must be used to inform the visualization of these layers. We also hope to release WikAPIdia in the near future as an open source project, although the need for it has decreased in the past year thanks to the development of other excellent Wikipedia APIs ([3], [19] and [42]). However, WikAPIdia has certain unique features that the Wikipedia research community may find helpful, particularly with regard to its full support of the "snippet" concept.

**ACKNOWLEDGMENTS**
We are grateful to the Deutsche Telekom Laboratories for partially funding this research.

*Note: This version of the paper contains a fix for a reference issue that appeared in the original version.*